\newcommand{\ud}{\rm d}
\newcommand{\un}{~\mathrm}
\newcommand{\mum}{~\mu\mathrm{m}}

\documentclass[aps,prl,reprint,superscriptaddress]{revtex4-1}

\usepackage[T1]{fontenc}
\usepackage{color}
\usepackage{psfrag}
\usepackage[pdftex]{graphicx}
\usepackage{dcolumn}
\usepackage{bm}
\usepackage{amsmath} 
\usepackage[pdftex]{graphicx}
\usepackage{epstopdf}

\begin{document}
\title{Fluctuations of global energy release and crackling in nominally brittle heterogeneous fracture}

\author{J. Barés}
\altaffiliation[Present address: ]{Duke University, Durham, North Carolina 27708, USA}
\affiliation{Laboratoire SPHYNX, Service de Physique de l'Etat Condens{\'e}, IRAMIS, CEA Saclay, CNRS URA 2464, 91191 Gif-sur-Yvette, France}
\author{M.~L. Hattali}
\altaffiliation[Present address: ]{Univ Paris-Sud, UPMC Univ Paris 6, CNRS, UMR 7608, Laboratoire FAST, Bat 502 -- Campus Universitaire, F-91405 Orsay, France}
\affiliation{Laboratoire SPHYNX, Service de Physique de l'Etat Condens{\'e}, IRAMIS, CEA Saclay, CNRS URA 2464, 91191 Gif-sur-Yvette, France}
\author{D. Dalmas}
\affiliation{Unité Mixte CNRS/Saint-Gobain, Surface du Verre et Interfaces, 39 Quai Lucien Lefranc, 93303 Aubervilliers cedex, France}
\author{D. Bonamy}
\email{Daniel.Bonamy@cea.fr}
\affiliation{Laboratoire SPHYNX, Service de Physique de l'Etat Condens{\'e}, IRAMIS, CEA Saclay, CNRS URA 2464, 91191 Gif-sur-Yvette, France}


\begin{abstract}
The temporal evolution of mechanical energy and spatially-averaged crack speed are both monitored in slowly fracturing artificial rocks. Both signals display an irregular burst-like dynamics, with power-law distributed fluctuations spanning a broad range of scales. Yet, the elastic power released at each time step is proportional to the global velocity all along the process, which enables defining a material-constant fracture energy. We characterize the intermittent dynamics by computing the burst statistics. This latter displays the scale-free features signature of crackling dynamics, in qualitative but not quantitative agreement with the depinning interface models derived for fracture problems. The possible sources of discrepancies are pointed out and discussed. 
\end{abstract}

\maketitle

Predicting when and how solids break continues to pose significant fundamental challenges \cite{Bonamy11_pr,Alava06_ap}. This problem is classically addressed within the framework of continuum mechanics, which links deterministically the degradation of a solid to the applied loading. Such an idealization, however, fails in several situations. In heterogeneous solids upon slowly increasing loading for instance, the fracturing processes are sometimes observed to be erratic, with random events of sudden energy release spanning a variety of scales. Such dynamics are e.g. revealed by the acoustic emission accompanying the failure of various materials \cite{Petri94_prl,Garcimartin97_prl,Davidsen07_prl,Baro13_prl} and, at much larger scale, by the seismic activity going along with earthquakes \cite{Bak02_prl,Corral04_prl}; A generic observation in this field is the existence of scale-free statistics for the event energy \cite{Bonamy09_jpd}. 

These avalanche dynamics \cite{Sethna01_nature} have attracted much recent attention. They were originally thought to be inherent to quasi-brittle fracture, where the solid starts by accumulating diffuse damage through microfracturing events before collapsing when a macroscopic crack percolates throughout the microcrack cloud \cite{vanMier12_book}. Phenomenological models such as fiber bundle models (see \cite{Pradhan10_rmp} for review) or random fuse models (see \cite{Alava06_ap} for review) developed in this case reproduce qualitatively the avalanche dynamics with a minimal set of ingredients. More recently, it has been demonstrated \cite{Bonamy08_prl} that a situation of nominally brittle fracture, involving the destabilization and propagation of a single crack, can also yield erratic dynamics. Within the linear elastic fracture mechanics (LEFM) framework, the in-plane motion of a crack front was mapped to the problem of a long-range (LR) elastic interface propagating within a two-dimensional (2D) random potential \cite{Schmittbuhl95_prl,Ramanathan97_prl}, so that the driving force self-adjusts around the depinning threshold \cite{Bonamy08_prl}. This approach reproduces, in a simplified 2D configuration,  the local and irregular avalanches evidenced in the space-time dynamics of an interfacial crack growing along a weak heterogeneous plane \cite{Maloy06_prl}. There exists theoretical arguments to extend this approach to the bulk fracture of real three-dimensional (3D) solids and crackling dynamics at the global (specimen) scale are anticipated \cite{Ponson10_ijf,Bares13_prl}. Still, fracture experiments are crucially missing to demonstrate this point.
  
The study reported here aims at filling this gap. Fracture experiments in heterogeneous solids made of sintered polymer beads are found to display irregular burst-like dynamics at the global scale, with large, power-law distributed, fluctuations for the mean failure speed $v(t)$ and overall mechanical energy $E(t)$ stored in the specimen. Yet and despite their individual giant fluctuations, the ratio between $v(t)$ and the power release $\ud E(t)/ \ud t$ remains constant and defines a continuum-level scale material-constant fracture energy. The burst statistics displays the scale-free features predicted in elastic interface models. Still, the agreement remains qualitative only, and the scaling exponents are different from those predicted. The possible sources of discrepancies are discussed.

{\em Experiments --} The experiments were carried out on artificial rocks made of sintered polystyrene beads: i) a mold filled by monodisperse polystyrene beads was heated to $T=105~^\circ\mathrm{C}$ ($90\%$ of the temperature at glass transition) and compressed (pressure $p=4.2\un{MPa}$) between the two jaws of an Instron electromechanical machine while keeping $T=105~^\circ\mathrm{C}$; ii) the mold was then unloaded and slowly cooled down to ambient temperature and the obtained sample was extracted from it. 
This sintering process provides heterogeneous solids with homogeneous microstructures, the length-scale of which is set by the bead diameter $d$. In all the experiments reported here, $d=500~\mu\mathrm{m}$. Large enough heterogeneity scales, indeed, is requested to observe global crackling at finite driving rate \cite{Bares13_prl} and the fracture of sintered materials with smaller $d$ ($250\,\mu$m, $140\,\mu$m, $80\,\mu$m and $40\,\mu$m) were observed to display continuum-like dynamics.  

\begin{figure}
\begin{center}
\includegraphics[width=\columnwidth]{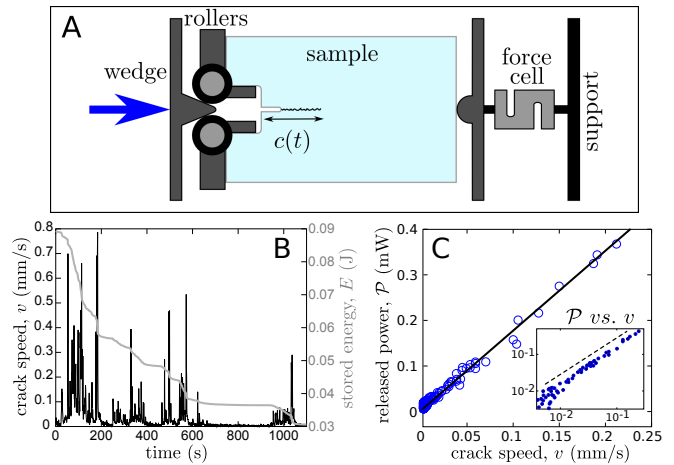}
\caption{A: Sketch of the Wedge-Splitting test. B: Zoomed view of the crack tip speed $v(t)$ (black) and stored mechanical energy $E(t)$ (gray) as a function of time in a typical fracture experiment (Here, $V_{wedge}=16\un{nm/s}$). C: Instantaneous power release $\mathcal{P}=-\ud E/\ud t$ as a function of $v(t)$ for all $t$. The axes are linear in the main panel, and logarithmic in the inset. In both panels, straight lines indicate proportionality. The proportionality constant $\Gamma=100\pm10\un{J/m}^2$ gives the material's fracture energy. In both panels B and C, the coarse-graining time is $\delta t=0.2\un{s}$}
\label{fig1}
\end{center}
\end{figure}

In the so-obtained materials, stable cracks were driven by means of wedge splitting fracture tests (see Fig. \ref{fig1}A and Refs. \cite{Scheibert10_prl,Guerra12_pnas} for details): Parallelepiped samples of size $140 \times 125 \times{15} \un{mm}$ in the $x$ (propagation), $y$ (loading) and $z$ (sample thickness) were loaded in mode I by pushing a wedge at constant speed $V_{wedge}$ into a $25 \times 25\un{mm}$ cut out on one the two $(y-z)$ edges. An initial seed crack ($10\un{mm}$-long) was introduced with a razor blade in the middle of the cut. It prevents dynamic fracture and enables growing slow stable cracks. Two go-between steel blocks were placed between the wedge and the specimen to limit parasitic mechanical dissipation and ensure the damage and failure processes to be the sole dissipation source for mechanical energy in the system (see \cite{Bares13_phd} for details). 

The wedge speed $V_{wedge}$ was varied from $16\un{nm/s}$ to $1.6\mum/s$. During each test, the force $f(t)$ applied by the wedge was monitored in real time by a S-type Vishay cell force (acquisition rate of $50\un{kHz}$, accuracy of $1\un{N}$). As soon as the wedge starts to push on the specimen (time origin set at this onset), $f$ increases. When $f$ gets large enough ($\sim 200-300\un{N}$), the seed crack starts to propagate. This propagation was imaged at the specimen surface via a camera (USB2 uEye from IDS Imaging Development, space and time accuracy of $130~\mu\mathrm{m}$ and $0.1\un{s}$, respectively), providing the instantaneous length $c_{surface}(t)$ of the crack edge at the surface. The instantaneous mechanical energy $E(t)$ stored in the specimen is given by $E(t)=\frac{1}{2} f(t)\times V_{wedge} \times t$, and the instantaneous {\em mean} crack length $c(t)$ (i.e. spatially averaged over specimen thickness) is obtained from the knowledge of the instantaneous specimen stiffness $k(t)=f(t)/V_{wedge}\times t$. Indeed, in a linear elastic material, the curve $k~vs.~c$ is a continuous decreasing function set by the specimen geometry only, and independent of the other experimental parameters (e.g. $V_{wedge}$, microstructure parameters...). We hence measured the curves  $k~vs.~c_{surface}$ in each of our experiments, averaged them over all our experiments, and smoothed the result via a Tikhonov regularization. The so-obtained curve defines the curve $k~vs.~c$ for our fracture geometry; it was checked this curve is identical to that obtained using 2D finite element calculations (software Castem 2007) on the exact experimental geometry, assuming plane stress conditions. This reference curve $k(c)$ was used to infer $c(t)=k^{-1}(f(t)/V_{wedge} \times t)$ from the signal $f(t)$ [$k^{-1}(x)$ denoting the inverse function of $k(x)$]. Time derivation of $c(t)$ finally provides the instantaneous crack speed $v(t)$.

{\em Results --} Figure \ref{fig1}(A) presents the time evolution of $v(t)$ and $E(t)$ in a typical fracture experiment. These profiles exhibit the intermittent features characteristic of crackling dynamics, with random violent bursts (resp. sudden drops) in $v(t)$ (resp. in $E(t)$). The superposition of the two also reveals that the velocity bursts coincide with the energy drops. Beyond this occurrence coincidence, the fluctuation amplitude $v(t)$ is proportional to the power $\mathcal{P}(t)=-\ud E /\ud t$ released at each moment $t$ (fig. \ref{fig1}(B)). This proportionality was observed in all our experiments, irrespectively of the wedge speed. It betrays the characteristics of a nominally brittle fracture: $\mathcal{P}(t) = G(t) \times v(t)$ where $G(t)=-\ud E /\ud c$ is the energy release rate. Now, for a stable crack slowly driven in a nominally brittle material, LEFM states that $G(t) \sim \Gamma$  where the fracture energy $\Gamma$ is a material constant. In other words, a nominally brittle fracture compatible with LEFM assumptions yields $\mathcal{P}(t) = \Gamma \times v(t)$ at all times $t$, irrespectively of the precise values of $\mathcal{P}(t)$ and $v(t)$, as observed here. In this scenario, the proportionality constant in fig. \ref{fig1}(B) gives $\Gamma$ for the considered material: Here $\Gamma=100\pm 10\un{J/m}^2$.      

\begin{figure}
\begin{center}
\includegraphics[width=\columnwidth]{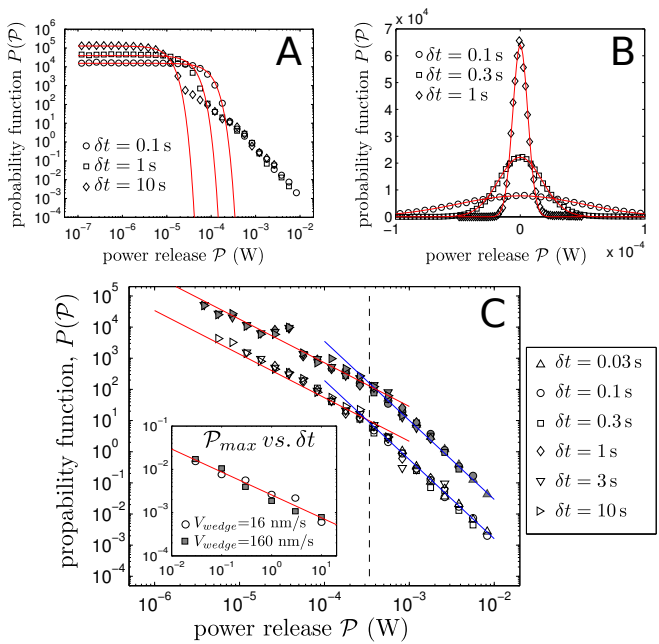}
\caption{Distribution of instantaneous power release for different coarse-graining time $\delta t$ plotted in logarithmic (panel A) and linear scales (panel B). Here, $V_{wedge}=16\un{nm/s}$. Plain curves are Gaussian distributions with zero average and a variance $\sigma(\delta t)$ prescribed so as the Gaussian curve fits the experimental data on the small scale plateau. C, main panel: Same graphs as in A after having withdrawn the noise-dominated Gaussian part of each distribution. Empty and filled symbols correspond to $V_{wedge}=16\un{nm/s}$ and $V_{wedge}=160\un{nm/s}$, respectively. The curves associated to $V_{wedge}=160\un{nm/s}$ have been shifted vertically for sake of clarity. Distributions involve 3094 events for $V_{wedge}=16\un{nm/s}$ and 5299 events for $V_{wedge}=160\un{nm/s}$. Red and blue straight lines are power-law fits with exponents $a_{small}=1.4\pm0.15$ (small scale regime) and $a_{large}=2.5\pm0.1$ (large scale regime). Vertical dash line locates the crossover $\mathcal{P}_C \approx 0.34\un{mW}$. C, inset: maximum value $\mathcal{P}_\mathrm{max}$ observed for $\mathcal{P}(t)$ as a function of $\delta t$. Red line shows the $1/\sqrt{\delta t}$ dependency.}
\label{fig2}
\end{center}
\end{figure}

We turn now to the statistical characterization of the crack dynamics. We analyzed the temporal evolution $\mathcal{P}(t)$ in preference to that of $v(t)$ since the former is directly obtained from the experimental measurement of applied force $f(t)$, while the latter calls for the addition of the $k~vs.~c$ curve. The distribution of instantaneous power released is first analyzed. Note that, in experiments, an "instantaneous" quantity is actually averaged over a finite time scale $\delta t$, the value of which affects the fluctuation amplitude.  The distributions of $\mathcal{P}(t)$, hence, have been computed for different values of $\delta t$. A Gaussian distribution (centered at zeros, standard deviation decreasing as $1/\sqrt{\delta t}$) is observed at small scales (plain lines in Figs. \ref{fig2}A and B), and a power-law tail is observed at large scales (Fig. B). The Gaussian part at small $\mathcal{P}$ is observed throughout the whole experiments, even in the preliminary loading phases where cracks do not propagate. It results from the noise in the measurement of the force signal. This noise yields a $\delta t$-dependent resolution limit $\mathcal{P}_r$ below which the true fracture-induced fluctuations of $\mathcal{P}(t)$ cannot be deconvoluted from the Gaussian noise. Conversely, the probability that a fluctuation of size $\mathcal{P}(t) > \mathcal{P}_r$ is due to noise rapidly is insignificant. Keeping only the relevant part above $\mathcal{P}_r$, the distributions all collapse onto a single master curve (Fig. \ref{fig2}C) exhibiting two power-law scaling, a small-scale regime with a scaling exponent $a_{small}=1.4\pm0.15$ and a large-scale regime with $a_{large}=2.5\pm 0.1$. The two scaling regimes, together with the value of the associated crossover ($\mathcal{P}_C \approx 0.34\un{mW}$), depend neither on $\delta t$, nor on the loading rate $V_{wedge}$. Conversely, the maximal value $\mathcal{P}_\mathrm{max}$ (resp. $v_\mathrm{max}$) of $\mathcal{P}(t)$ (resp. $v(t)$) decreases with $\delta t$, as $1/\sqrt{\delta t}$ (Inset in Fig. \ref{fig2}C), as expected for independent fluctuation peaks. Note that the large scale power-law exponent $a_{large}=2.5\pm 0.1$ observed here at the global scale for fracture experiments in bulk solids is very close to that reported on the local velocity fluctuations in 2D situations of interfacial cracks, both experimentally \cite{Maloy06_prl,Tallakstad13_prl} and numerically \cite{Gjerden14_frontiers}. This supports the conjecture that at large scales, brittle 3D fracture can be reduced to a 2D elastic interface problem \cite{Bonamy08_prl,Bonamy09_jpd,Bares13_prl}. Conversely, the small scale power-law regime with $a_{small}=1.4\pm 0.1$ observed here differs from that observed in the 2D interfacial configuration.

\begin{figure}
\begin{center}
\includegraphics[width=\columnwidth]{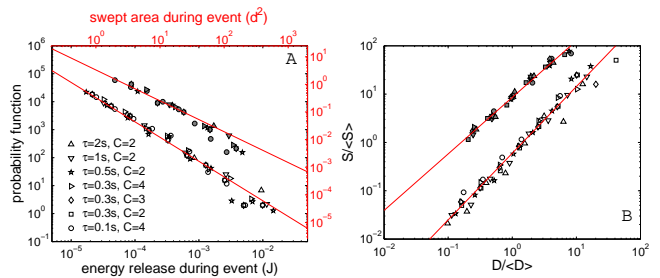}
\caption{A: Distribution of the avalanche size $S$, defined either as the energy released (black/bottom) or as the area swept (red/top) during the event. In the latter case, the area is normalized by $d^2$ where $d=500\,\mu$m is the grain diameter B: Scaling between normalized size $S/\langle S \rangle$ and duration $D/\langle D \rangle$. In both panels, the various symbols correspond to various values for $\{\delta t,C\}$ (specified in the inset of A) and $V_{wedge}$ (empty symbol for $V_{wedge}=16\un{nm/s}$, filled ones for $V_{wedge}=160\un{nm/s}$; the latter have been shifted vertically for sake of clarity). Analyzes involve 899 avalanches for $V_{wedge}=16\un{nm/s}$ and 473 events for $V_{wedge}=160\un{nm/s}$. Straight lines are power-law fits $P(S) \propto S^{-\tau}$ and $S \propto D^{\gamma}$, with $\tau=1.4 \pm 0.1$ (resp. $\tau=1.1 \pm 0.15$) and $\gamma=1.38 \pm 0.05$ (resp. $\gamma=1.17 \pm 0.05$) for $V_{wedge}=16\un{nm/s}$ (resp. $V_{wedge}=160\un{nm/s}$).}
\label{fig3}
\end{center}
\end{figure}

The scale-free statistics observed for the fluctuations $\mathcal{P}(t)$ (or equivalently for the fluctuations $v(t)$) is a first hint toward crackling dynamics. We adopt the standard procedure in the field, and identify the underlying avalanches with the bursts where $\mathcal{P}(t)$ is above a prescribed reference level $\mathcal{P}_r=C\langle \mathcal{P} \rangle$. Then, the avalanche duration $D$ of each pulse is given by the interval between the two intersections of $\mathcal{P}(t)$ with $\mathcal{P}_r$, and the avalanche size $S$ is defined as the energy released during the event, i.e. the integral of $\mathcal{P}(t)$ between the two intersection points. Note that, in conventional elastic interface formalism, the avalanche size $S$ is expressed as the total area $A$ swept by the front between two successive pinned configurations. The two definitions are equivalent since, in the nominally brittle situation experienced here, an event releasing an energy $S$ creates fracture surfaces of area $A=S/\Gamma$. As expected for a crackling signal, $S$ follows a power-law distribution $P(S) \propto S^{-\tau}$ over nearly three order of magnitude, with event areas up to $\sim 400 $ times the elementary one $d^2$ (Fig. \ref{fig3}:A). [a break in the scaling, around $3\times 10^{-3}\un{J}$ and $10^{-3}\un{J}$ for $V_{wedge}=16\un{nm/s}$ and $V_{wedge}=160\un{nm/s}$ cannot be precluded. Also, the mean avalanche size goes as a power-law with $D$, $S\propto D^{\gamma}$ (Fig. \ref{fig3}:B). The exponents $\tau$ and $\gamma$ are independent of the prescribed values for $\delta t$ and $C$. Conversely, they both decrease with the loading rate, from $\{\tau=1.4 \pm 0.1, \gamma=1.38 \pm 0.05\}$ at $V_{wedge}=16\un{nm/s}$ to $\{\tau=1.1 \pm 0.15, \gamma=1.17 \pm 0.05\}$ at $V_{wedge}=160\un{nm/s}$.

\begin{figure}
\begin{center}
\includegraphics[width=\columnwidth]{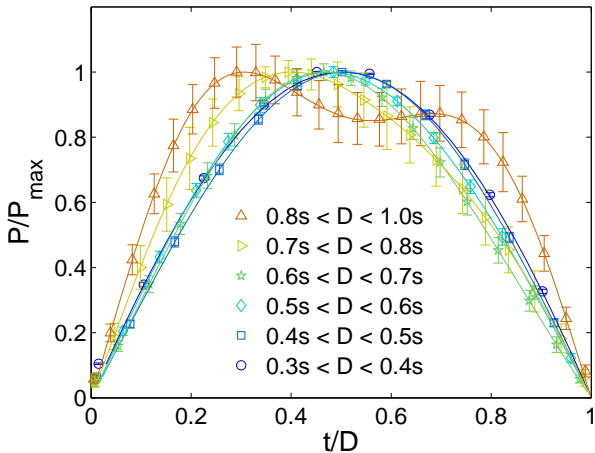}
\caption{Temporal avalanche shape for different avalanche durations $D$ ($V_{wedge}=16\un{nm/s}$). Note the shape flattening and leftward asymmetry that develop as $D$ increases.}
\label{fig4}
\end{center}
\end{figure}

To complete the dynamics characterization, we computed the average temporal avalanche shape. This observable, indeed, provides an accurate characterization of the considered crackling signal and has been measured in a variety of systems \cite{Zapperi05_natphys,Mehta06_pre,Laurson06_pre,Papanikolaou11_natphys,Danku13_prl,Laurson13_natcom}. The standard procedure was adopted here: First, we identified all the pulses $i$ of duration $D_i$ falling within a prescribed interval $[D_{min},D_{max}]$); and second, we averaged the shape $\mathcal{P}^i(t)/\mathcal{P}^i_{max}~vs.~t/D_i$ over all the collected pulses $i$. Figure \ref{fig4}:A shows the resulting shape and its evolution with $D$. Two observations should be noted: i) The shape is parabolic for small $D$ and flattens as $D$ increases; ii) A small, but clear leftward asymmetry is observed as $D$ increases: The bursts start faster than they stop. Quantitatively, the shape evolution (how fast it flattens and how large the asymmetry is) depends on $V_{wedge}$.

A similar shape flattening was observed in Barkhausen experiments \cite{Papanikolaou11_natphys}. Therein, it was shown to result from the finite value of the demagnetizing factor $k$. Such a shape flattening is thus expected in the LR interface model for crack growth \cite{Bonamy08_prl}, where the unloading factor plays the same role as $k$ \cite{Bares13_prl}. Conversely, the numerical simulation of this LR interface model yielded {\em symmetrical} shapes \cite{Bares13_phd} (at finite driving rate) or slight rightward asymmetry \cite{Laurson13_natcom} (at vanishing driving rate). In Barkhausen experiments, in contrast, a leftward asymmetry was observed, and attributed to the eddy currents, which provides a negative effective mass to the domain walls\cite{Zapperi05_natphys}. We conjecture that the viscoelastic nature of the polymer rocks fractured here acts in a similar way by providing a negative inertia to the crack front (i.e. the addition of a retardation term in the LR interface model of crack \cite{Bonamy08_prl}).  

{\em Concluding discussion --} The experiments reported here demonstrate that crackling at global scale can be observed in nominally brittle fractures (due to the propagation of a single crack) and is not restricted to quasi-brittle (multi-fracturing) situation: Three main observations emerge: i) Despite their individual giant fluctuations, the ratio between spatially-averaged velocity and power release remains fairly constant and defines a continuum-level scale material-constant fracture energy; ii) The event size, defined either as the increase of crack length or as the energy release during the event, is power-law distributed, and scales as a power-law with the event duration; iii) the associated exponents depend on the crack loading rate. 

These observations are in qualitative agreement with what is predicted by a recent model \cite{Bonamy08_prl} identifying stable crack growth with a LR elastic interface driven in a random potential so that the driving force self-adjust around the depinning threshold. Still, the agreement is qualitative only: i) In the universality class of the LR depinning transition,  the scaling exponents are predicted to be $a_{small}=0.38$ \cite{Dobrinevski13_phd}, $\tau=1.28$ \cite{Bonamy09_jpd}, and $\gamma=1.80$ \cite{Bonamy09_jpd}, significantly different from the experimental values measured here; ii) These predicted exponents are independent of the driving rate \cite{Bares14_frontiers}, contrary to what is observed here. This discrepancy is thought to result from the finite width of the fracture specimens; the LR elastic kernel in the interface model \cite{Bonamy08_prl} arises from Rice's perturbative analysis of the elastic problem of a corrugated crack front embedded in a sample of infinite width \cite{Rice85_jam}. Conversely, it is interesting to note that the variations of $\tau$ and $\gamma$ observed in our experiments are compatible with those expected in the mean-field ABBM model \cite{Alessandro90_jap}: In the ABBM model, i) $\tau$ decreases from 3/2 to 1 as the driving rate increases; ii) $\gamma$ exhibits two values: $\gamma=2$ for short pulses, $\gamma=1$ for large ones, with a crossover decreasing with driving rate \cite{Papanikolaou11_natphys}. In our experiments, the mean-field approximation may be relevant since the separation between the microstructure scale (bead size $d=500~\mu\mathrm{m}$) and the continuum-level scale (specimen width: $15\un{mm}=30d$) is quite small. The simplicity of the ABBM model has allowed the derivation of exact analytical solutions for the avalanche distribution and shape \cite{Papanikolaou11_natphys,Colaiori08_ap,Dobrinevski12_pre,Dobrinevski13_pre}. A very interesting future extension of this study is to accurately characterize how the avalanche statistics evolves with the driving rate and to confront those against the ABBM solutions. Work in this direction is under progress.        

\begin{acknowledgments}
Support through ANR project MEPHYSTAR (ANR-09-SYSC-006-01) is gratefully acknowledged. Special thanks to Thierry Bernard for technical support. 
\end{acknowledgments}

%


\end{document}